\documentclass{article}

\usepackage[english]{babel}
\usepackage{subcaption} 

\usepackage[letterpaper,top=2cm,bottom=2cm,left=3cm,right=3cm,marginparwidth=1.75cm]{geometry}

\usepackage{amsmath}
\usepackage{graphicx}
\usepackage[style=apa, backend=biber]{biblatex}
\usepackage[colorlinks=true, allcolors=blue]{hyperref}

\input{glyphtounicode}
\usepackage[T1]{fontenc}

\usepackage[utf8]{inputenc} 
\usepackage{lmodern}        

\title{AI-Agents for Culturally Diverse Online Higher Education Environments }
\author{Fuze Sun, Paul Craig, Lingyu Li, Shixiangyue Meng, Chuxi Nan}

\begin{document}
\maketitle

\begin{abstract}
As the global reach of online higher education continues to grow, universities are increasingly accommodating students from diverse cultural backgrounds \parencite{tereshko2024culturally}. This can present a number of challenges including linguistic barriers \parencite{ullah2021linguistic}, cultural differences in learning style \parencite{omidvar2012cultural}, cultural sensitivity in course design \parencite{nguyen2022cultural} and perceived isolation when students feel their perspectives or experiences are not reflected or valued in the learning environment \parencite{hansen2022belonging}. Ensuring active engagement and reasonable learning outcomes in such a environments requires distance educational systems that are not only adaptive but also culturally resonant \parencite{dalle2024cultural}. Both embodied and virtual AI-Agents have great potential in this regard as they can facilitate personalized learning and adapt their interactions and content delivery to align with students' cultural context. In addition Generative AI (GAI), such as, Large Language Models (LLMs) can amplify the potential for these culturally aware AI agents to address educational challenges due to their advanced capacity for understanding and generating contextually relevant content \parencite{wang2024large}. This chapter reviews existing research and suggests the usage of culturally aware AI-Agents, powered by GAI, to foster engagement and improve learning outcomes in culturally diverse online higher education environments.
\end{abstract}


\section{Introduction}

Online degree programs are provided by many universities around the world offering an increasing number of learners more flexible, convenient and cost-effective access to higher education  \parencite{diaz2021foreign}. This form of higher education is currently undergoing a radical evolution as advancements in artificial intelligence (AI), such as the emergence of Large Language Model, reshape traditional online learning methodologies \parencite{wang2024large}. 

\begin{figure}[ht]
    \centering
    \includegraphics[width=0.8\textwidth]{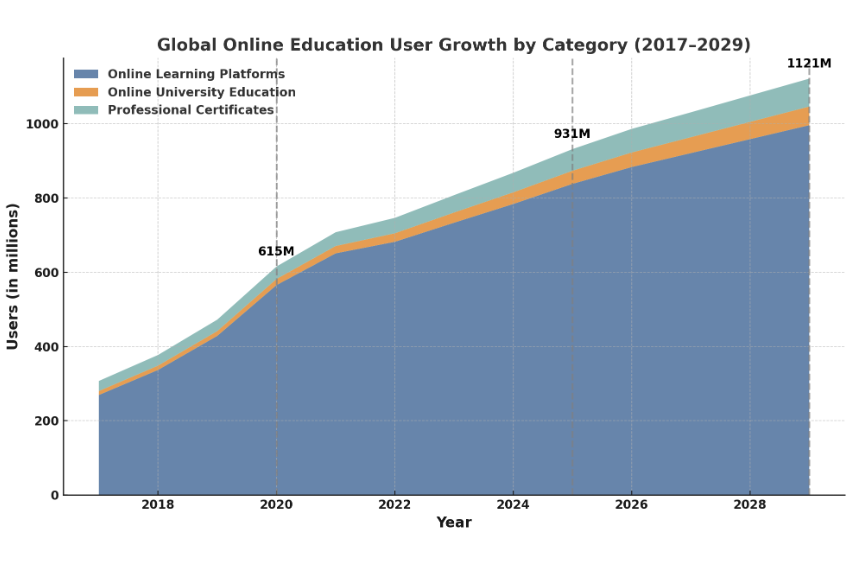}
    \caption{Global Online Education User Growth and Prediction by Category (2017–2029).
This figure illustrates the expansion of online education users worldwide, with steady growth across professional certificates, online learning platforms, and online university education. The diversification of educational formats highlights how online education is evolving to meet varied learner needs, contributing to greater inclusivity and cultural diversity. Data source: (\cite{statista2025}).}
    \label{fig:chart1}
\end{figure}

\begin{figure}[ht]
    \centering
    \includegraphics[width=0.8\textwidth]{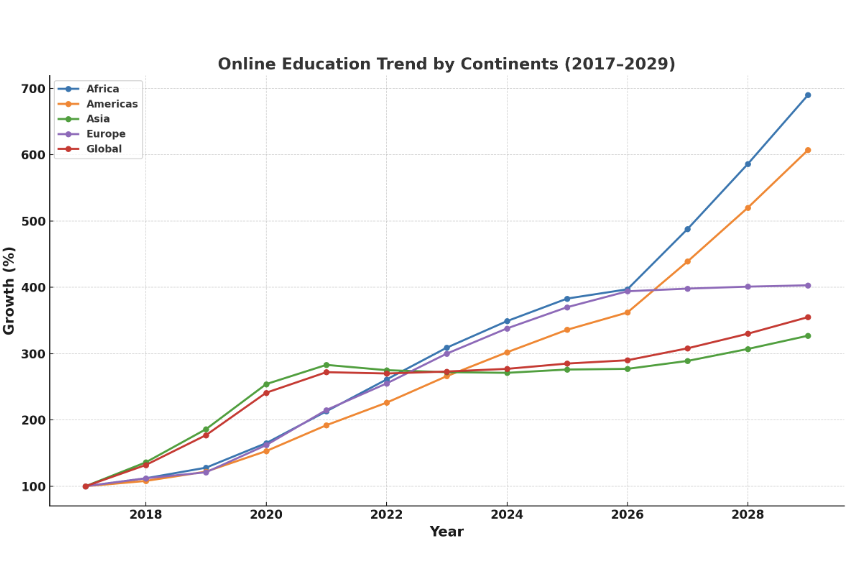}
    \caption{Online Education Trends by Continents (2017–2029).
This figure shows the rise of online education across continents, with especially rapid growth in Africa and the Americas. The increasing adoption of online learning globally reflects not only quantitative growth but also a broadening cultural landscape, as diverse populations participate in online higher education. Data source: (\cite{statista2025}).}
    \label{fig:chart2}
\end{figure}

The growing diversity of the student population has increased the demand for culturally responsive teaching practices \parencite{lawrence2020teaching}. As illustrated in Figures ~\ref{fig:chart1} and ~\ref{fig:chart2}, the global growth in online education is not only quantitative but also culturally expansive. The rapid increase in users across continents, especially in  Africa and the America, along with the diversification of educational formats, indicates a significant rise in cultural diversity among online learners. Despite this, student engagement and motivation remain critical factors in determining academic success \parencite{akpen2024engagement} and conventional online teaching approaches (notwithstanding their advantages in affordability and accessibility) often struggle to provide the sort of engaging and adaptive learning experiences that it would be possible to deliver with the more traditional on-campus face-to-face education \parencite{haron2021challenges}. As a result, online education often falls short in delivering the sort of dynamic, culturally resonant, and personalized experiences that would be expected from an onsite lecture with an educator who is physically present.

Inclusion and diversity are equally central to AI-powered learning. \textcite{XIA2022104582} propose that self-determination theory (SDT) provides a framework for engaging underrepresented learners in AI education. Through meeting students' psychological needs for competence, relatedness and autonomy, teachers can mitigate achievement gaps, enabling lower-achieving students to feel more confident and internally motivated. Such principles can create equitable learning environments that respond to heterogeneous needs across student populations when implemented in AI-tutoring systems. \textcite{inbook} further suggest that AI can enhance mental health support and accessibility in higher education. They emphasize that adaptive AI can personalize learning for students with disabilities and those facing psychological challenges. Inclusive design principles ensure that AI supports equal access, therefore addressing systemic inequities that often disadvantage marginalized groups.

In today’s diverse educational landscape, the demand for interactive and student-centered distance education depends not only on technological innovation \parencite{kerimbayev2023student}, but also on involving cultural narratives based on projective narrative that resonate with student’s identities and lived experience \parencite{al2025educational}. By leveraging these approaches, educators can cultivate a learning environment that is inclusive \parencite{bond2020engagement} and reflective of students' unique cultural perspectives \parencite{godsk2025engaging}. This has the potential to improve student engagement, enrich the learning experience, and empower students to connect their academic pursuits with their individual and collective identities. 

\begin{figure}[ht]
    \centering
    \begin{subfigure}{0.45\textwidth}
        \centering
        \includegraphics[width=\linewidth]{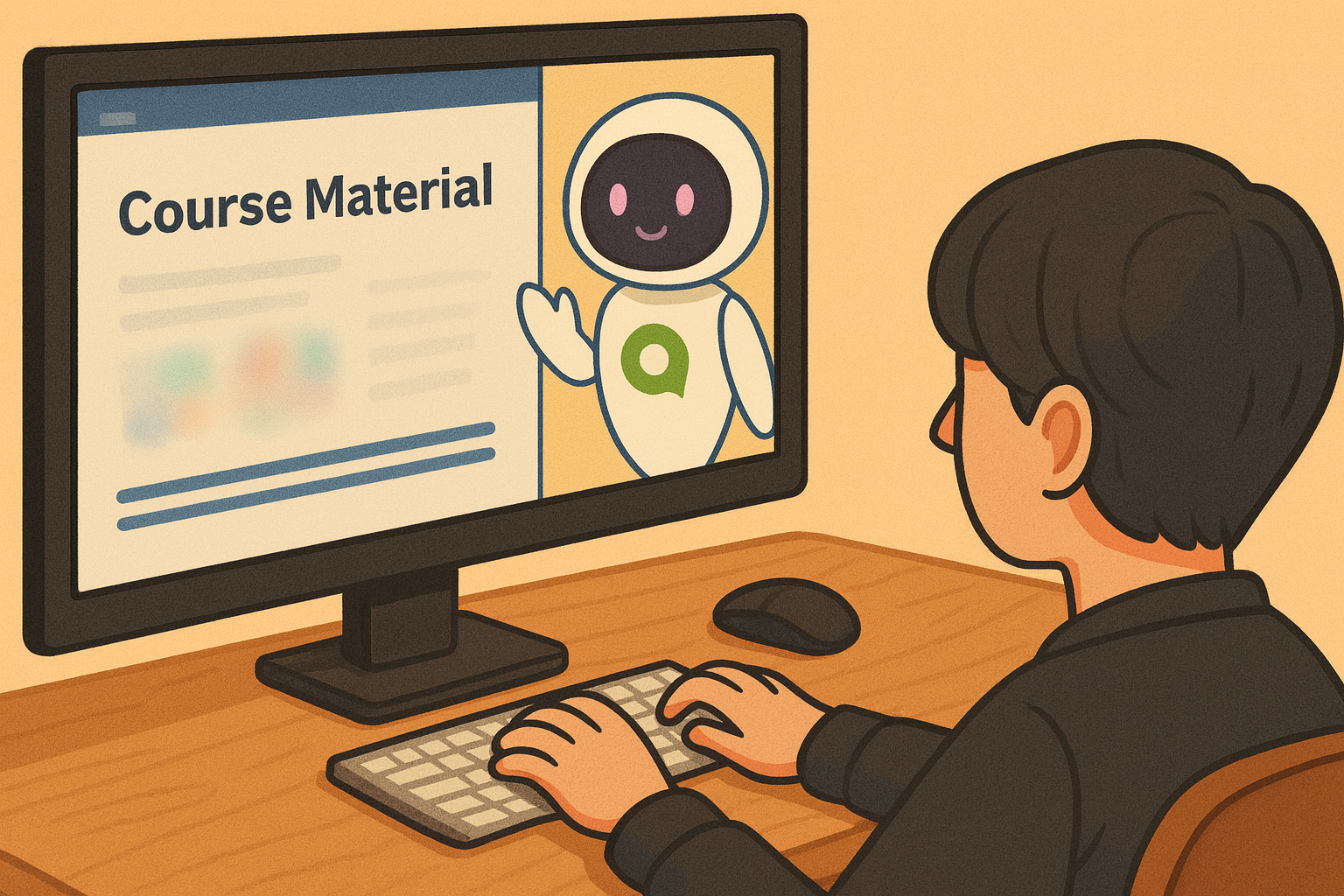}
        \caption{Virtual Tutor in Online Education}
        \label{fig:virtual_tutor}
    \end{subfigure}
    \hfill
    \begin{subfigure}{0.45\textwidth}
        \centering
        \includegraphics[width=\linewidth]{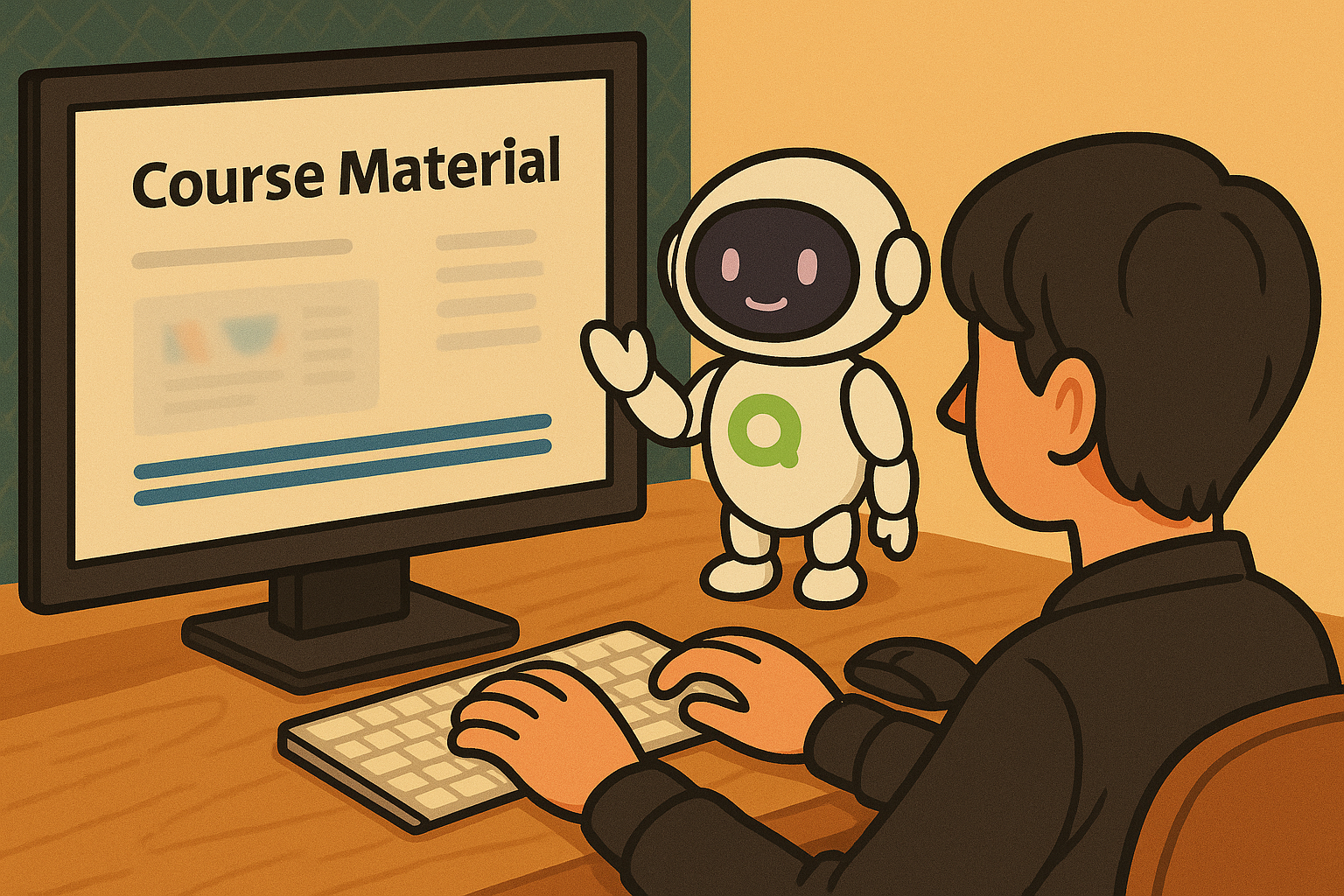}
        \caption{Robot Tutor in Online Education}
        \label{fig:robot_tutor}
    \end{subfigure}
    \caption{Comparison of Virtual and Robot Tutors in Online Education}
    \label{fig:comparison_tutors}
\end{figure}

This chapter discusses the potential of GAI powered virtual \& embodied agents to facilitate cultural awareness and support cultural narratives in an online higher education background by considering to major components that would be combined to realize such an agent. These are the Large-Language Model, the system for emotional recognition and responsiveness, the presence and physicality of the agent, and the memory architecture for personalization. We also consider the two main types of AI-Agent, which are virtual and robotic tutors (see Figure~\ref{fig:comparison_tutors}), as well as the challenges faced by developers and educators in this direction.

\section{Generative AI and Large Language Models (LLMs) with regards to Culture}

Generative AI has significantly transformed the field of AI research, enabling us to use computers in mimicking human creativity to generate natural text content, images, audio, and even videos. Large Language Models are at the core of this technology using advanced deep learning models trained on vast datasets to generate coherent and contextually relevant content. These models, such as OpenAI’s GPT, Meta’s LLaMA and Google’s Gemini, leverage a deep learning architecture called a transformer which is based on the multi-head attention mechanism to jointly attending multiple positions, for processing and generating natural language \parencite{liu2021multi}. Their ability to understand and generate context-aware responses makes LLMs particularly effective for various applications, including AI-Agent powered virtual or physically embodied tutors targeted for assisting online higher education programs. 

Emerging evidence suggests that cultural factors profoundly affect the effectiveness and adoption of AI systems. \textcite{cannavale2025} show that cultural dimensions such as uncertainty avoidance, risk aversion, and collectivism shape national strategies for AI implementation. These findings indicate that AI systems designed without sensitivity to cultural values risk reinforcing digital divides. In educational setting, this highlights the importance of equipping AI tutors with the capability that not only personalize content delivery but also align with diverse cultural expectations of authority, trust, and collaboration. This cultural dimension is particularly relevant for higher education, where cross-cultural and international classrooms are the norm. Such alignment ensures that learning technologies build trust across varied cultural groups and avoid reproducing exclusionary biases embedding in the training data.

An AI-Agent framework combined with GAI (which will be introduced in the following sections) can help them maximize the ability of GAI to produce human-like cultural resonant interactions. However, the effectiveness of these models depends on what we can think of as their \textit{intelligence level}. This includes the number of parameters (the numerical values within the model that define its behavior and capabilities), the quality and quantity of training data used, the techniques used in training, and the sophistication of the model architecture. This level can move the system beyond mere text-based interaction to multi-modal processing of a student’s expressions and behaviors.

Most implemented AI tutoring systems primarily based on unimodal Large Language Models (LLMs), which can only process text inputs and outputs \parencite{chen2023tutoring}. Although these models were effective in providing customized explanations and answering student queries, their interaction channel was limited to textual communication, making it difficult to show empathy for fully engaging students or catering to cultural sensitivity. Text-based LLMs often fall short in satisfying diverse learning preferences and cultural backgrounds (\cite{guizani2025education}). Moreover, these systems lack the ability to interpret multi-sensory cues, such as student’s facial expressions or tone of voice, which restrict their ability to adapt to individual behaviors, support long-term retention, or enhance user’s motivation in culturally diverse educational settings \parencite{cornelio2021multisensory}. 

\begin{figure}[ht]
    \centering
    \includegraphics[width=0.8\textwidth]{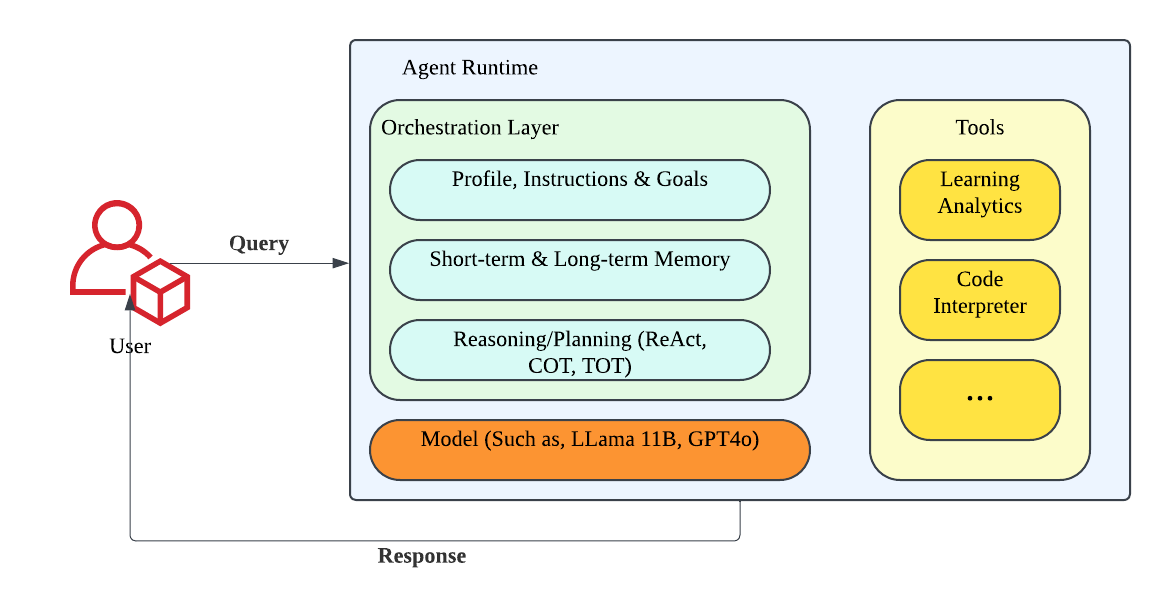}
    \caption{Student interacts with Multi-Modal AI-Agent and its inner components.}
    \label{fig:Interactive_process}
\end{figure}

AI-Agents powered by multi-modal LLMs such as Vision-Language model (see Figure~\ref{fig:Interactive_process}) represent the next stage in AI-driven education, integrating multiple sensory channels-text, speech, and vision, to provide a more interactive and empathetic learning environment. This shift is particularly critical in distance higher education, where personalized and culturally sensitive interaction is essential for engagement. Multi-modal AI-Agents enhance learning experience by enabling virtual or robot tutors to do the following:

\begin{itemize}
    \item Process and respond to multiple input types (e.g., understanding and answer a student's verbal question while recognizing his emotional state by extracting the facial expression that can facilitate empathetic interaction) 
    \item Generate multi-modal responses, including spoken explanations, interactive visual demonstrations through images, and real-time feedback customized based on student’s cultural background and learning history stored within the system’s memory module. 
    \item Dynamically adapting to student engagement levels by analyzing tone of voice, facial expression, and non-verbal gestures, enabling culturally responsive tutoring that builds rapport and enhances personalization. 
\end{itemize}

These adaptive features provided by Generative AI resonate with the importance of cultural narratives in education, as students from different backgrounds may express engagement, confusion, or satisfaction in diverse ways. The ability to interpret and respond to these nuanced cues is essential for creating inclusive and culturally responsive learning environments \parencite{henriksen2025creativity}. Recent advancements in multi-modal AI models, such as closed-source GPT-4o and open-source LLama 3.3, have paved the way for the development of cost-efficient AI-Tutors capable of understanding and empathetically interacting with students in a more natural and engaging manner.  No matter wether they are implemented as virtual avatars or embodied robotics tutors, these AI-Agent powered systems have the potential to bring breakthrough to distance higher education by providing adaptive, context-aware, and empathetic learning experiences that go beyond the limitations of traditional text-based rigid AI tutors. 

\section{Emotion and Empathy in Culturally Responsive Virtual and Robotic Tutors}

Emotions play a critical role in the learning process, influencing student’s engagement, motivation, and cognitive processing \parencite{sun2025integratingemotionalintelligencememory}. In the scenario of AI-Agent powered virtual and robotics tutors for assisting distance higher education, integrating customized LLMs which enable emotional intelligence to achieve empathy, can significantly enhance the learning experience. Situational meanings and individual interpretations can elicit emotional responses  \parencite{schutz2006reflections} with these interpretations often being shaped by cultural narratives and norms. It is therefore crucial for AI tutors to recognize, interpret, and adapt to student’s emotional states while remaining sensitive to their cultural backgrounds and learning context. This culturally attuned empathy can significantly enhance learning experience, fostering deeper engagement and motivation. 

Research has revealed that emotion-aware AI-Agent systems, such as those incorporating the OCC emotional model, can recognize, interpret and effectively respond to human emotions, therefore enhancing engagement, rapport, and satisfaction among users \parencite{kwon2007emotion}. This interpretation and responding to student's emotions requires detailed analysis of a key series of cues, like facial expressions \parencite{rawal2022facial} which are also essential considerations when the agent generates its own response to students in empathetic manner.  However, understanding and responding to emotions is inherently culturally contextual. The interpretation of emotions, whether through facial expressions or gestures, vary across cultures. An AI-Tutor that fails to account for these cultural differences risks misinterpreting student’s emotional cues, leading to inadequate responses. To enable culturally-aware interactions, AI-Tutors, whether virtual or embodied, can be designed to infer student’s emotional states using data from cameras, microphones, and memory module. This methodology helps the AI-Agent to adapt its responses, fostering empathy through two key mechanisms. These are as follows. 

\begin{itemize}
    \item \textbf{Parallel Empathy:} Mirror the learner’s emotional state to create a sense of shared understanding. For instance, if a student appears enthusiastic, the tutor can respond with encouraging feedback and supportive gestures, reinforcing the positive emotion.
    \item \textbf{Reactive Empathy:} The AI-Agent responds appropriately to the learner’s emotions, offering tailored supports. For instance, if a student displays signs of frustration, the tutor can provide encouragements or culturally relevant examples that resonate with the student’s background and experiences. 
\end{itemize}

In interaction design, empathy is defined as the agent’s ability to recognize a user’s emotional state, thoughts, and context, then generate affective or cognitive responses to bring positive effects on human user's perception \parencite{park2022empathy}.  This capability is essential for both virtual and embodied robot tutors.  Whether conveyed through verbal tone, sentiment adaption in text, or culturally attuned gestures, empathetic responses help create emotionally supportive learning environments. In culturally diverse educational settings these empathetic interactions are even more significant, as student’s emotional expressions and communication preferences are shaped by their cultural backgrounds and experiences. 

Studies on socially interactive agents have demonstrated that learners perceive the experience better when AI tutors provide positive reinforcement and encouragement, particularly when they detect frustration or confusion of the participant \parencite{brown2014positive}. Since the growing importance of emotional intelligence in AI tutoring systems, it is critical to explore innovative approaches to integrate emotional recognition and responses methodologies into both virtual AI tutors and embodied robot tutors, such as implementing educational AI-Agent powered by customized vision-language model targeted for teaching. By doing so, online higher education can utilize these AI-driven tutors to create more engaging, empathetic, and effective learning environments, mitigating the lack of interaction inherent to this form of learning. 

\section{Non-Verbal Communication in Virtual and Robotic Tutors}

Non-verbal communication, such as gestures, eye movement, and facial expressions, plays an important role in enhancing the naturalness and intuitiveness of the interaction between students and AI-Agent powered virtual or robot tutors. These social cues are essential for creating culturally sensitive, empathetic, and supportive educational environments that foster stronger student engagement and improved learning outcomes. In worldwide diverse online educational contexts, where the expressions of engagement and understanding may vary, the design of non-verbal cues in AI systems becomes even more critical . 

Research has revealed that eye-contact significantly contributes to establish attentiveness and connection during the interaction process, which is the key to manage turn-taking in conversations and creating a supportive learning environment \parencite{admoni2017social}. However, the interpretation of eye-contact is culturally nuanced, while it signifies engagement in some cultures, it may be perceived as intrusive or disrespectful in others. Therefore, either virtual or embodied tutors must be designed to adapt their non-verbal behaviors based on cultural expectations. 

Virtual tutors can directly mimic human-like eye movements on the screen, ensuring culturally appropriate levels of eye contact \parencite{gorlatova2022eyesyn}. Similarly, some robot tutor also uses a screen as the face to show expressions, others which have a physical face with LEDs as eyes use head movements to disguise as eyes are moving. For both virtual and robot tutors, maintaining appropriate eye contact or mimicking attentive head movements can show active listening, making students feel acknowledged and understood. Analogously, non-verbal cues like nodding, facial expressions, and postures have been shown to maintain interest and engagement, since they convey emotional awareness for empathy \parencite{rodriguez2015bellboy}. These social cues help students perceive the tutor’s anthropomorphism, fostering a more natural and interactive communication dynamic. 

AI tutors can also extract and implement non-verbal cues to better interpret the mental states and intentions of students, enabling bidirectional non-verbal communication \parencite{fiore2013toward}. For example, detecting and recognizing when a student appears confused or frustrated through their facial expressions can prompt the robot tutor to provide clarification or encouragement. This kind of empathetic adaptive technique has massive potential for maintaining student engagement and facilitating deeper interaction. Another important factor in non-verbal communication is Gesture, which plays a critical role in reinforcing verbal explanations and convey empathy. Effective robot gestures for education could include the following.

\begin{itemize}
    \item A robot tutor can use hand gestures to show the size of a tangible object or the importance of a historical event, enhancing the student’s comprehension. 
    \item A virtual tutor on the screen can implement avatar animations, such as pointing, nodding, or leaning forward, to emphasize attentiveness and empathy. 
\end{itemize}

\begin{figure}[ht]
    \centering
    \includegraphics[width=0.65\textwidth]{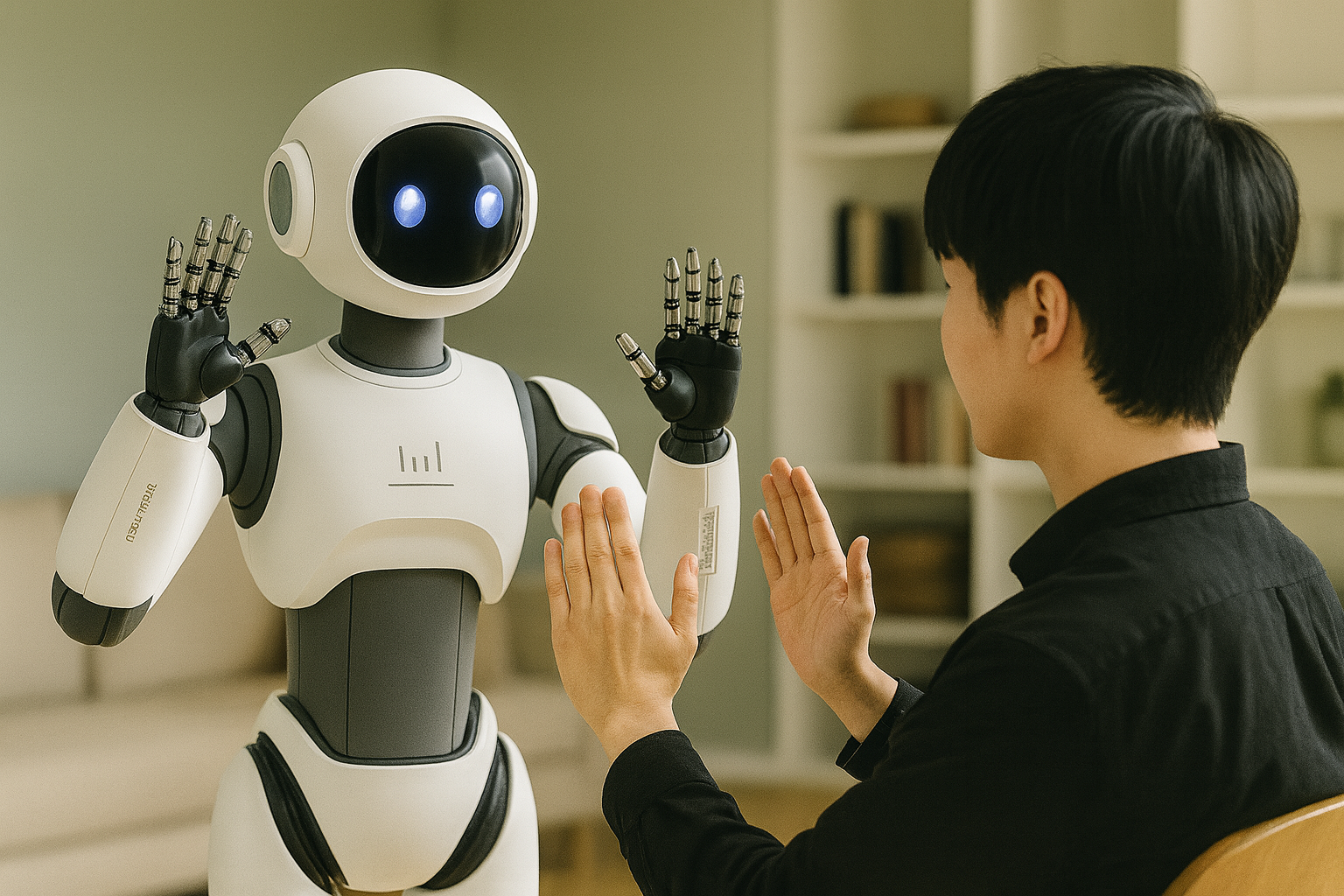}
    \caption{An assumption robot used to illustrate how robotic arms and gestures can serve as communicative components in HRI-based tutoring systems.}
    \label{fig:assumption_robot}
\end{figure}

As a core component of Human Robot Interaction(HRI), the robotic arm serves as a crucial medium for conveying gestures, allowing robot tutors to physically demonstrate and reinforce concepts in ways that mimic human communication. The robot issues action primitives such as pointing, grasping, and  co-manipulation, adapting these actions in response to student participation and feedback \parencite{10.1007/978-3-031-93539-8_16}. To support diverse students and instructional contexts, this process requires a top-down and explainable pipeline from perception to execution \parencite{chen2023precision}. Within such a pipeline, key interaction parameters, including separation distance, motion speed, confirmation frequency, and handover mode, can be programmed to maintain safety while accommodating cultural expectations and communication preferences. Figure~\ref{fig:assumption_robot} illustrates an assumption robot designed for such interaction.

\textcite{sidner2005explorations} have shown that learners direct more attention to the educational agents that employ gestures to indicate empathy, with feedback suggesting that these interactions are perceived more natural and meaningful compared to those lacking social cues. In educational scenarios, non-verbal cues such as index fingers gestures, eye gaze, or forward leaning can enhance the building of rapport, shaping the AI-Agent's role as a supportive tutor \parencite{kennedy2017impact}. However, it is important to recognize that non-verbal communication is culturally constructed, and the interpretation of gestures or expressions can vary significantly across cultures. For example, hand gestures that reflect agreement or encouragement in one culture may incur different or even negative connections in another. A simple gesture like a thumbs-up may be seen as positive in some contexts but perceived inappropriate or offensive in others. Therefore, AI tutors must be designed to adapt their non-verbal behaviors in ways that respect and reflect the cultural narratives and communication norms of diverse learners.

\section{Personalization and Cultural Memory in GenAI Tutors}

For online learning, personalization is known to be effective for enhancing student engagement, motivation and learning outcome \parencite{bernacki2021personalized}. Personalization is not only about adapting to a student’s academic needs but also about recognizing their culture backgrounds, learning preferences, and individual experiences. One of the key enablers that help to achieve personalization in AI-powered virtual and embodied tutors is the integration of memory architecture inside the AI-Agent that allows for remembering and adapting to individual student’s preferences, cultural narratives, and learning journeys over time. 

\begin{figure}[ht]
    \centering
    \includegraphics[width=0.8\textwidth]{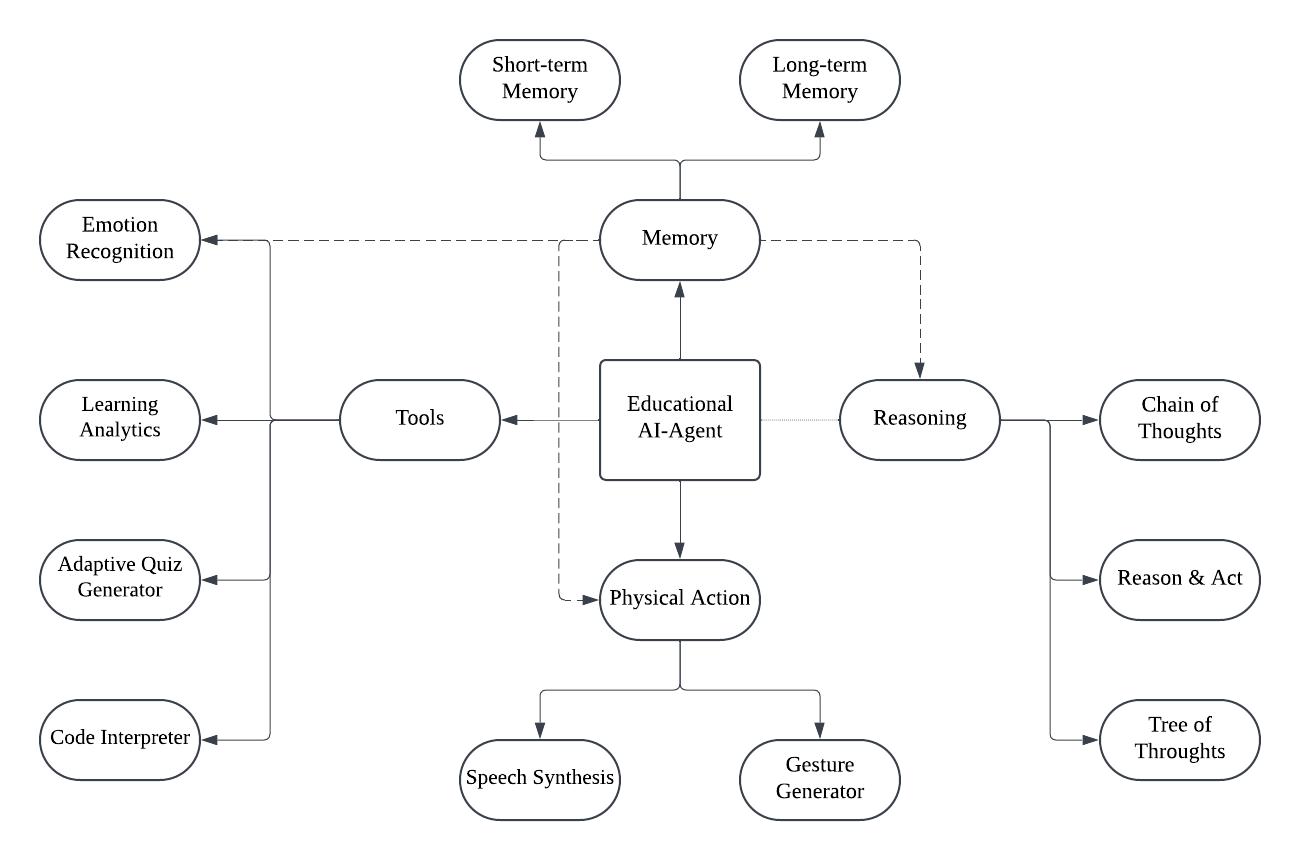}
    \caption{The memory module will coordinate with other parts (such as, Tool, Action) to achieve personalized learning experience.}
    \label{fig:memory}
\end{figure}

For embodied robot tutors, a memory module enables the robot to recall previous interaction records and individual student’s profile, facilitating more empathetic and customized conversations \parencite{ligthart2022memory}. The memory module is able to affect other parts of the agent to achieve personalized and empathetic actions (see Figure~\ref{fig:memory}), for example, the robot tutor that can remember a student’s preferred learning style, prior knowledge, or previous challenges will offer more targeted guidance and culturally relevant encouragement by notifying the Action and Tool module to act correspondingly for enhancing rapport and trust. Similarly, in virtual AI tutors, memory systems can support the development of culturally sensitive, adaptive interactions delivered by the avatar on the screen. This kind of memory-driven personalization methodology also contributes to mimicking human-like interactions, where recalling and adapting based on past experiences fosters a sense of being understood and valued \parencite{leyzberg2018effect}. In culturally diverse educational settings, this ability to recall and integrate culturally meaningful details is essential for fostering inclusive and respectful learning experiences.

A key component of effective memory architecture in AI-powered tutors is the integration of two types of memory, short-term and long-term memory. These can be described as follows.

\textbf{Memory Types in AI Tutors:}
\begin{itemize}
    \item Short-term memory focuses on processing information within the current interaction session, such as remembering recent conversation contents, detected emotional cues. This ensures that interactions feel immediate and contextually relevant, encouraging more natural and engaging dialogues. 
    \item Long-term memory is about storing and recalling information from past sessions or structured data collections, such as a student’s academic process, learning preferences, or culturally relevant experiences. This enables deeper and more structured personalization over time, fostering stronger rapport and trust by demonstrating that the AI tutor “remembers” and values the student’s unique learning journey. 
\end{itemize}

\begin{figure}[ht]
    \centering
    \includegraphics[width=0.8\textwidth]{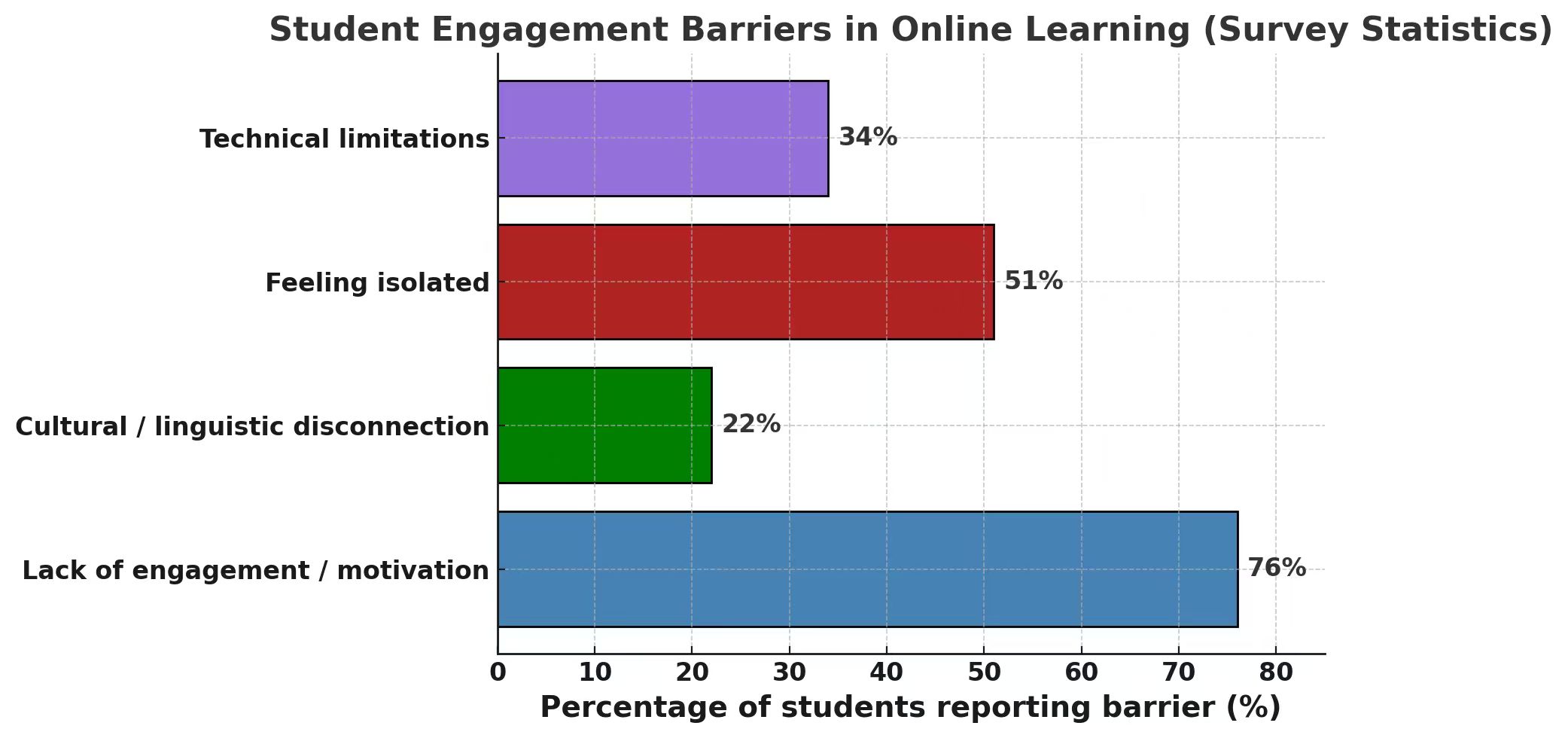}
    \caption{Student Engagement Barriers in Online Learning. Survey-based statistics indicate that the most significant barrier reported by students is the lack of engagement or motivation (76\%), followed by feelings of isolation (51\%), technical limitations (34\%), and cultural/linguistic disconnection (22\%). Data sources: \parencite{pew2021, bevens2024, zheng2015, nasfaa2020}.}
    \label{fig:barrier}
\end{figure}

From a psychological perspective, personalized learning contributes to enhancing situational interest which is a short-term engagement factor, and it is crucial for developing long-term interest in a subject \parencite{reber2018personalized}. In culturally diverse educational settings, personalization becomes even more significant, as student’s engagement is shaped by their culture backgrounds, learning preferences, and lived experience. An AI tutor that can remember and adapt to individual student needs fosters a stronger sense of connection, belonging, and inclusivity, particularly in distance learning scenarios where real-time interaction is limited. Recent emerging technologies in the field of LLMs have significantly improved the potential for personalized AI-Agent. For example, the LangChain framework enables the integration of memory modules into the AI-driven applications, supporting both short-term interaction processing and long-term information storage. This dual memory structure allows AI tutors to maintain context over extended interactions, ensuring that responses are not only personalized but also culturally sensitive and contextually appropriate. This kind of AI-Agent implemented in educational scenario can better address the last three barriers in online learning which is spotted during the review of previous studies (see Figure~\ref{fig:barrier}).

A potential disadvantage of this is that storing sensitive student data could result in data breaches or unauthorized access \parencite{bernacki2021personalized}. Additionally, the memory module might inadvertently amplify biases present in the training data, leading to culturally insensitive or unfair responses \parencite{guizani2025education}. Furthermore, implementing and maintaining these systems is resource-intensive, which can limit scalability and increase costs \parencite{fattahi2024advancing}. Addressing these challenges is crucial to ensure that memory modules effectively enhance learning while mitigating potential drawbacks.

Despite these challenges, integrating a memory module into an educational AI-Agent's framework that powers virtual and embodied AI tutors has the potential to create an adaptive, culturally aware, and empathetic learning experiences for students enrolled in distance higher education. Through referencing individual student’s past records in the system, AI tutors can deliver personalized support that improves engagement, motivation, and academic success. What is more, this approach fosters deeper, more meaningful educational relationships, ensuring that students feel recognized, respected, and included throughout their learning journal. Such personalization attuned culturally is important for addressing the diverse needs of learners in global, distance education settings, contributing to the broader goal of creating accessible and culturally responsive educational platforms. 

\section{Pedagogical Trade-offs: Virtual vs. Embodied Tutors in Cultural Contexts}

Virtual tutors are AI-driven avatars integrated within digital learning platforms shown up on the screen, interacting with students through text, speech, or visual interfaces. Their primary advantages lie in their affordability and accessibility, since they can be easily invoked as long as there is a stable network connection and educational platforms always provide them for free. Virtual tutor's model can also be shaped by prompt word engineering or fine-tuned by corresponding data sets to provide empathetic responses \parencite{chen2025empathyagent}, utilizing the advancements in Large Language Models (LLMs) to tailor feedback and interaction styles according to individual learners’ cultural backgrounds and preferences. This helps them to be particularly effective in online higher education platforms, where enrolled students are from diverse cultural and linguistic contexts. In addition, virtual tutors are easier to update and integrate into online education platform, allowing for flexible content delivery and real-time interaction. 

The limitation of virtual tutors is their lack of physical presence and non-verbal engagement, which can hinder emotional connection and rapport-building.  Although virtual tutors can simulate empathy through text, speech or gestures on the screen, they struggle to replicate the richness of non-verbal cues, such as physical gestures, facial expressions. These are particularly valued in cultures where face-to-face interaction and physical presence are proven to be significantly more effective for building trust and engagement.

\begin{figure}[ht]
\centering
\includegraphics[width=0.8\textwidth]{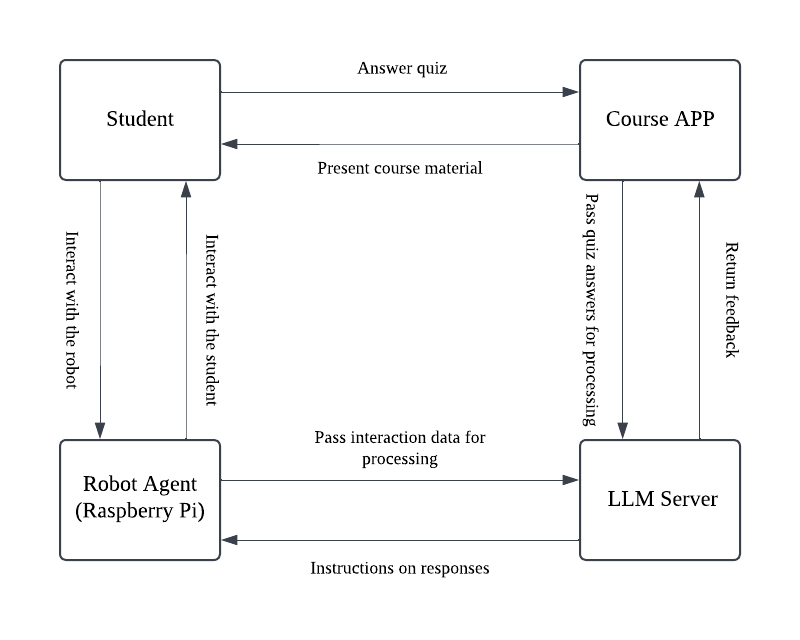}
\caption{The key components of the robot tutor integrated education.}
\label{fig:robot_tutor_environment}
\end{figure}

In contrast to virtual agents, embodied robot tutors act as tangible, human-like avatars to learning environments, providing deeper emotional engagement through multi-modal interactions, such as speech, gestures, and facial expressions \parencite{aylett2025embodied}. Their physical body can mimic authentic social interactions, enhancing rapport and motivations, especially for culturally diverse contexts where non-verbal communication is highly valued. Embodied tutors can convey empathy through culturally relevant gestures and postures, enhancing inclusive learning experiences that resonate with diverse students. The learning environment of the robot tutor integrated education is shown below (Figure \ref{fig:robot_tutor_environment}).

\begin{figure}[ht]
    \centering
    \includegraphics[width=0.5\textwidth]{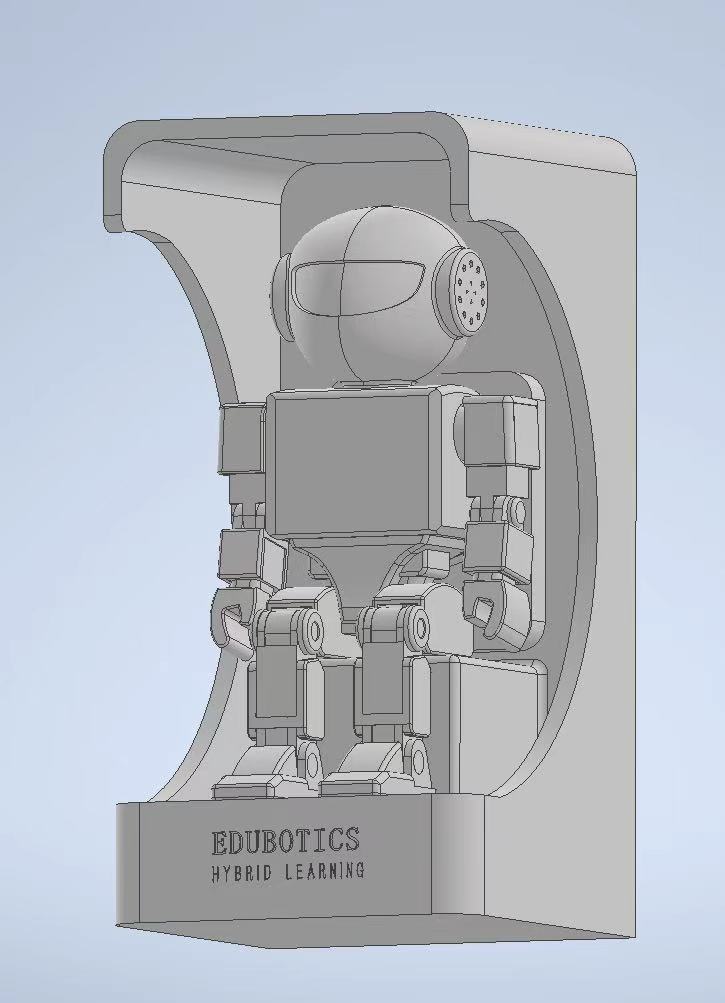}
    \caption{Blueprint of a 3D-printed Robot}
    \label{fig:blue_print}
\end{figure}

Robot tutors also come with some significant challenges for the educator and the learner. On the education provider side, embodied systems are often resource-intensive, with high cost for development, such as ensuring the robot’s gestures and expressions are culturally appropriate is complex, as social cues vary across culture. From the learner’s point of view, the high initial cost of purchasing a robot for home use can be prohibitive, particularly for students from under-resourced backgrounds. However, more and more open-source, 3D printed robot projects are published online which can decrease the cost of owning a robot agent(see Figure~\ref{fig:blue_print}). Besides the price, additional expenses related to software updates, maintenance and repairs further increase the burden for the learner. In addition, issues of compatibility and interoperability may arise, as robotic tutors purchased from one educational provider may not work well with the systems and software of other platforms. 

\section{Challenges, Future Directions, and Conclusion}

Generative AI, including both text-based and multimodal models, has revolutionized intelligent, empathetic tutoring by enabling rich, context-aware interactions through text, speech, visuals, and even gestures. Unlike traditional rule-based educational software, these AI systems are equipped with human-like attributes such as emotional intelligence, memory architectures, and gesture controller. This allows them to dynamically adapt to students' needs, recognize engagement levels using emotional and cultural cues, and provide personalized learning feedback and pathways. Personalized feedback on course assignments fosters a sense of community and strengthens relationships with learners, ultimately enhancing the quality of interactions in online courses \parencite{song2023enhancing}. This supports the primary goal of multicultural education: to reform the educational experience so that students from diverse racial, ethnic, and social-class backgrounds can access equitable learning opportunities \parencite{banks2015dimensions}. These findings have paved the way for AI-powered, culturally-aware tutors that can operate in both virtual and robot forms, each offering distinct advantages in transforming online higher education.

Emerging virtual tutors, powered by customized AI agents that utilize Large Language Models (LLMs) as their intelligent core, currently operate on digital learning platforms such as Xpert from Edx to provide students with on-demand, interactive,b and personalized support. Research shows that embedding human-like emotional and psychological elements in tutoring can significantly enhance engagement and learning outcomes \parencite{limo2023personalized}. Leveraging frameworks such as LangChain, these tutors can provide real-time feedback, guide coursework, and insert culturally related content to promote inclusion within digital learning systems. In order to better results, embodied robot tutors extend personalization by utilizing a human-like presence through multi-modal interactions such as speech synchronized with physical gestures. Positioned alongside the student, the AI-driven robot promotes attentiveness, motivation, and collaborative learning by simulating culturally resonated social interactions.

In addition to these advances, serval challenges remain. Privacy concerns surrounding the usage of personal data, the difficulty of scaling emotional intelligence, and the need for authentic cultural adaptability keep posing significant hurdles, together with the persistent risk of embedding cultural biases from training datasets. Nevertheless, Generative AI demonstrates substantial potential to transform online higher education by enabling empathetic, adaptive, and culturally resonated learning experiences. Virtual tutors offer flexibility and scalability, while embodied robot tutors promote motivation and engagement through their presence and multi-modal capabilities. In the future, research and development should prioritize robust ethical frameworks, enhanced memory architectures, and culturally adaptive designs that coordinate advanced cultural intelligence, emotional responsiveness, and memory-driven personalization into both virtual and robotic AI-Agents, paving the way for truly empathetic and inclusive educational systems capable of meeting the diverse needs of global learners.

\printbibliography

@article{tereshko2024culturally,
  title={Culturally Diverse Student Engagement in Online Higher Education: A Review},
  author={Tereshko, Lisa M and Weiss, Mary Jane and Cross, Samantha and Neang, Linda},
  journal={Journal of Behavioral Education},
  pages={1--33},
  year={2024},
  publisher={Springer}
}

@article{hansen2022belonging,
  title={College Students’ Belonging and Loneliness in the Context of Remote Online Classes during the COVID-19 Pandemic},
  author={Hansen-Brown, Ashley A. and Sullivan, Sean and Jacobson, Brianna and Holt, Blake and Donovan, Shaelyn},
  journal={Online Learning},
  volume={26},
  number={4},
  pages={323--346},
  year={2022},
  publisher={Online Learning Consortium}
}

@inbook{nguyen2022cultural,
  title={Designing for Cultural Responsiveness},
  author={Nguyen, Nhung},
  booktitle={Designing Learning Experiences for Inclusivity and Diversity},
  pages={1--15},
  year={2022},
  publisher={CAUL Open Educational Resources Collective}
}

@article{omidvar2012cultural,
  title={Cultural Variations in Learning and Learning Styles},
  author={Omidvar, Pegah and Tan, Bee Hoon},
  journal={Turkish Online Journal of Distance Education},
  volume={13},
  number={4},
  pages={269--278},
  year={2012},
  publisher={TOJDE}
}

@article{ullah2021linguistic,
  title={Linguistic Barriers in Online Teaching at Undergraduate Level in the University of Malakand Pakistan},
  author={Ullah, Zaheer and Ali, Shaukat and Hussain, Saddam},
  journal={Sir Syed Journal of Education \& Social Research},
  volume={4},
  number={1},
  pages={158--163},
  year={2021},
  publisher={SSJESR}
}

@article{dalle2024cultural,
  title={Cultural Dimensions of Technology Acceptance and Adaptation in Learning Environments},
  author={Dalle, Juhriyansyah and Aydin, Hasan and Wang, Charles Xiaoxue},
  journal={Journal of Formative Design in Learning},
  volume={8},
  pages={99--112},
  year={2024},
  publisher={Springer}
}

@article{wang2024large,
  title={Large Language Models for Education: A Survey and Outlook},
  author={Wang, Shen and Xu, Tianlong and Li, Hang and Zhang, Chaoli and Liang, Joleen and Tang, Jiliang and Yu, Philip S. and Wen, Qingsong},
  journal={arXiv},
  year={2024},
  url={http://arxiv.org/pdf/2403.18105v2}
}

@article{diaz2021foreign,
  title={FOREIGN STUDENTS'EXPERIENCE IN AN ONLINE INTERNATIONAL MARKET ANALYSIS COURSE IN MEXICO.},
  author={Diaz, Eduardo R},
  journal={Journal of Educators Online},
  volume={18},
  number={1},
  year={2021}
}

@article{lawrence2020teaching,
  title={Teaching as dialogue: Toward culturally responsive online pedagogy},
  author={Lawrence, April},
  journal={Journal of Online Learning Research},
  volume={6},
  number={1},
  pages={5--33},
  year={2020},
  publisher={Association for the Advancement of Computing in Education (AACE)}
}

@article{akpen2024engagement,
  title={Impact of online learning on student's performance and engagement: a systematic review},
  author={Akpen, Catherine Nabiem and Asaolu, Stephen and Atobatele, Sunday and Okagbue, Hilary and Sampson, Sidney},
  journal={Discover Education},
  volume={3},
  pages={205},
  year={2024},
  publisher={Springer}
}

@article{haron2021challenges,
  title={The Challenges and Constraints of Online Teaching and Learning in the New Normal Environment},
  author={Haron, Habibah @ Norehan and Masrom, Maslin and Ya’acob, Suraya and Sabri, Safiyyah Ahmad},
  journal={International Journal of Academic Research in Business and Social Sciences},
  volume={11},
  number={4},
  pages={1284--1295},
  year={2021},
  publisher={Human Resource Management Academic Research Society}
}

@article{kerimbayev2023student,
  title={A student-centered approach using modern technologies in distance learning: a systematic review of the literature},
  author={Kerimbayev, Nurassyl and Umirzakova, Zhanat and Shadiev, Rustam and Jotsov, Vladimir},
  journal={Smart Learning Environments},
  volume={10},
  pages={61},
  year={2023},
  publisher={Springer}
}

@article{al2025educational,
  title={Educational Technology and Its Impact on Learning},
  author={Al Tal, Safwat},
  booktitle={Technology for Societal Transformation},
  pages={29--44},
  year={2025},
  publisher={Springer}
}

@article{bond2020engagement,
  title={Mapping research in student engagement and educational technology in higher education: a systematic evidence map},
  author={Bond, Melissa and Buntins, Katja and Bedenlier, Svenja and Zawacki-Richter, Olaf and Kerres, Michael},
  journal={International Journal of Educational Technology in Higher Education},
  volume={17},
  pages={2},
  year={2020},
  publisher={Springer}
}

@article{liu2021multi,
  title={Multi-head or single-head? an empirical comparison for transformer training},
  author={Liu, Liyuan and Liu, Jialu and Han, Jiawei},
  journal={arXiv preprint arXiv:2106.09650},
  year={2021}
}

@article{godsk2025engaging,
  title={Engaging students in higher education with educational technology},
  author={Godsk, Mikkel and Møller, Karen Louise},
  journal={Education and Information Technologies},
  volume={30},
  pages={2941--2976},
  year={2025},
  publisher={Springer}
}

@article{fattahi2024advancing,
  title={Advancing education with large language models: a systematic review of potential, limitations, and business opportunities},
  author={FattahiBavandpour, Reza},
  journal={LUT University},
  year={2024},
  url={https://lutpub.lut.fi}
}

@article{chen2023tutoring,
  title={Empowering Private Tutoring by Chaining Large Language Models},
  author={Chen, Yulin and Ding, Ning and Zheng, Hai-Tao and Liu, Zhiyuan and Sun, Maosong and Zhou, Bowen},
  journal={arXiv},
  year={2023},
  url={https://arxiv.org/abs/2309.08112}
}

@article{guizani2025education,
  title={A systematic literature review to implement large language model in higher education: issues and solutions},
  author={Guizani, Sghaier and Mazhar, Tehseen and Shahzad, Tariq and Ahmad, Wasim and Bibi, Afsha and Hamam, Habib},
  journal={Discover Education},
  volume={4},
  pages={35},
  year={2025},
  publisher={Springer}
}

@article{cornelio2021multisensory,
  title={Multisensory Integration as per Technological Advances: A Review},
  author={Cornelio, Patricia and Velasco, Carlos and Obrist, Marianna},
  journal={Frontiers in Neuroscience},
  volume={15},
  pages={652611},
  year={2021},
  publisher={Frontiers Media SA},
  url={https://doi.org/10.3389/fnins.2021.652611}
}

@article{henriksen2025creativity,
  title={Generative AI, Creativity, Culture, and the Future of Learning},
  author={Henriksen, Danah and Oster, Nicole and Mishra, Punya and McCaleb, Lindsey},
  journal={TechTrends},
  volume={69},
  pages={3--9},
  year={2025},
  publisher={Springer},
  url={https://link.springer.com/article/10.1007/s11528-024-01036-y}
}

@article{gorlatova2022eyesyn,
  title={EyeSyn: Psychology-inspired Eye Movement Synthesis for Gaze-based Activity Recognition},
  author={Gorlatova, Maria and Lan, Guohao and Scargill, Tim},
  journal={International Conference on Information Processing in Sensor Networks},
  year={2022},
  publisher={IEEE},
  url={https://doi.org/10.1109/IPSN.2022.1234567}
}

@article{bernacki2021personalized,
  title={A Systematic Review of Research on Personalized Learning: Personalized by Whom, to What, How, and for What Purpose(s)?},
  author={Bernacki, Matthew L. and Greene, Meghan J. and Lobczowski, Nikki G.},
  journal={Educational Psychology Review},
  volume={33},
  pages={1675--1715},
  year={2021},
  publisher={Springer},
  url={https://link.springer.com/article/10.1007/s10648-021-09615-8}
}

@article{aylett2025embodied,
  title={An Embodied Empathic Tutor: Enhancing Emotional Engagement in Learning Environments},
  author={Aylett, Ruth and Jones, Aidan},
  journal={International Journal of Humanoid Robotics},
  volume={22},
  pages={45--67},
  year={2025},
  publisher={Springer},
  url={https://doi.org/10.1007/s12369-025-01234-x}
}

@article{schutz2006reflections,
  title={Reflections on investigating emotion in educational activity settings},
  author={Schutz, Paul A and Hong, Ji Y and Cross, Dionne I and Osbon, Jennifer N},
  journal={Educational psychology review},
  volume={18},
  pages={343--360},
  year={2006},
  publisher={Springer}
}

@inproceedings{kwon2007emotion,
  title={Emotion interaction system for a service robot},
  author={Kwon, Dong-Soo and Kwak, Yoon Keun and Park, Jong C and Chung, Myung Jin and Jee, Eun-Sook and Park, Kyung-Sook and Kim, Hyoung-Rock and Kim, Young-Min and Park, Jong-Chan and Kim, Eun Ho and others},
  booktitle={RO-MAN 2007-The 16th IEEE International Symposium on Robot and Human Interactive Communication},
  pages={351--356},
  year={2007},
  organization={IEEE}
}

@article{rawal2022facial,
  title={Facial emotion expressions in human--robot interaction: A survey},
  author={Rawal, Niyati and Stock-Homburg, Ruth Maria},
  journal={International Journal of Social Robotics},
  volume={14},
  number={7},
  pages={1583--1604},
  year={2022},
  publisher={Springer}
}

@article{park2022empathy,
  title={Empathy in human--robot interaction: Designing for social robots},
  author={Park, Sung and Whang, Mincheol},
  journal={International journal of environmental research and public health},
  volume={19},
  number={3},
  pages={1889},
  year={2022},
  publisher={MDPI}
}

@inproceedings{brown2014positive,
  title={The positive effects of verbal encouragement in mathematics education using a social robot},
  author={Brown, LaVonda N and Howard, Ayanna M},
  booktitle={2014 IEEE integrated STEM education conference},
  pages={1--5},
  year={2014},
  organization={IEEE}
}

@article{admoni2017social,
  title={Social eye gaze in human-robot interaction: a review},
  author={Admoni, Henny and Scassellati, Brian},
  journal={Journal of Human-Robot Interaction},
  volume={6},
  number={1},
  pages={25--63},
  year={2017},
  publisher={Journal of Human-Robot Interaction Steering Committee}
}

@article{rodriguez2015bellboy,
  title={A bellboy robot: Study of the effects of robot behaviour on user engagement and comfort},
  author={Rodriguez-Lizundia, Eduardo and Marcos, Samuel and Zalama, Eduardo and G{\'o}mez-Garc{\'\i}a-Bermejo, Jaime and Gordaliza, Alfonso},
  journal={International Journal of Human-Computer Studies},
  volume={82},
  pages={83--95},
  year={2015},
  publisher={Elsevier}
}

@article{fiore2013toward,
  title={Toward understanding social cues and signals in human--robot interaction: effects of robot gaze and proxemic behavior},
  author={Fiore, Stephen M and Wiltshire, Travis J and Lobato, Emilio JC and Jentsch, Florian G and Huang, Wesley H and Axelrod, Benjamin},
  journal={Frontiers in psychology},
  volume={4},
  pages={859},
  year={2013},
  publisher={Frontiers Media SA}
}

@article{sidner2005explorations,
  title={Explorations in engagement for humans and robots},
  author={Sidner, Candace L and Lee, Christopher and Kidd, Cory D and Lesh, Neal and Rich, Charles},
  journal={Artificial Intelligence},
  volume={166},
  number={1-2},
  pages={140--164},
  year={2005},
  publisher={Elsevier}
}

@article{kennedy2017impact,
  title={The impact of robot tutor nonverbal social behavior on child learning},
  author={Kennedy, James and Baxter, Paul and Belpaeme, Tony},
  journal={Frontiers in ICT},
  volume={4},
  pages={6},
  year={2017},
  publisher={Frontiers Media SA}
}

@inproceedings{ligthart2022memory,
  title={Memory-based personalization for fostering a long-term child-robot relationship},
  author={Ligthart, Mike EU and Neerincx, Mark A and Hindriks, Koen V},
  booktitle={2022 17th ACM/IEEE International Conference on Human-Robot Interaction (HRI)},
  pages={80--89},
  year={2022},
  organization={IEEE}
}

@article{leyzberg2018effect,
  title={The effect of personalization in longer-term robot tutoring},
  author={Leyzberg, Daniel and Ramachandran, Aditi and Scassellati, Brian},
  journal={ACM Transactions on Human-Robot Interaction (THRI)},
  volume={7},
  number={3},
  pages={1--19},
  year={2018},
  publisher={ACM New York, NY, USA}
}

@article{reber2018personalized,
  title={Personalized education to increase interest},
  author={Reber, Rolf and Canning, Elizabeth A and Harackiewicz, Judith M},
  journal={Current directions in psychological science},
  volume={27},
  number={6},
  pages={449--454},
  year={2018},
  publisher={Sage Publications Sage CA: Los Angeles, CA}
}

@article{song2023enhancing,
  title={Enhancing International Students' Participation in Online Courses through Culturally Responsive Pedagogy: A Systematic Literature Review},
  author={Song, Xiaoyan},
  year={2023}
}

@incollection{banks2015dimensions,
  title={The dimensions of multicultural education},
  author={Banks, James A},
  booktitle={Cultural Diversity and Education},
  pages={3--22},
  year={2015},
  publisher={Routledge}
}

@article{limo2023personalized,
  title={Personalized tutoring: ChatGPT as a virtual tutor for personalized learning experiences},
  author={Limo, Fernando Antonio Flores and Tiza, David Raul Hurtado and Roque, Maribel Mamani and Herrera, Edward Espinoza and Murillo, Jos{\'e} Patricio Mu{\~n}oz and Huallpa, Jorge Jinchu{\~n}a and Flores, Victor Andre Ariza and Castillo, Alejandro Guadalupe Rinc{\'o}n and Pe{\~n}a, Percy Fritz Puga and Carranza, Christian Paolo Martel and others},
  journal={Przestrze{\'n} Spo{\l}eczna (Social Space)},
  volume={23},
  number={1},
  pages={293--312},
  year={2023}
}

@article{chen2025empathyagent,
  author    = {Xinyan Chen and Jiaxin Ge and Hongming Dai and Qiang Zhou and Qiuxuan Feng and Jingtong Hu and Yizhou Wang and Jiaming Liu and Shanghang Zhang},
  title     = {EmpathyAgent: Can Embodied Agents Conduct Empathetic Actions?},
  journal   = {arXiv preprint},
  year      = {2025},
  volume    = {2503.16545},
  archivePrefix = {arXiv},
  primaryClass = {cs.CY},
  doi       = {10.48550/arXiv.2503.16545},
  url       = {https://doi.org/10.48550/arXiv.2503.16545}
}

@misc{statista2025,
  author       = {Statista},
  year         = {2025},
  title        = {Online education – Worldwide},
  url          = {https://www.statista.com/outlook/emo/online-education/worldwide#global-comparison},
  note         = {Accessed: 2025-08-27}
}

@misc{pew2021,
  author       = {{Pew Research Center}},
  year         = {2021},
  title        = {What we know about online learning and the homework gap amid the pandemic},
  howpublished = {\url{https://www.pewresearch.org/short-reads/2021/10/01/what-we-know-about-online-learning-and-the-homework-gap-amid-the-pandemic}},
  note         = {Accessed: 2025-08-28},
}

@article{bevens2024,
  author    = {Bevens, W. and Frontiers Editorial Office},
  year      = {2024},
  title     = {Loneliness, online learning and student outcomes},
  journal   = {Frontiers in Psychology},
  volume    = {15},
  pages     = {1408837},
  doi       = {10.3389/fpsyg.2024.1408837},
  url       = {https://doi.org/10.3389/fpsyg.2024.1408837}
}

@inproceedings{zheng2015,
  author    = {Zheng, S. and Rosson, M. B. and Shih, P. C. and Carroll, J. M.},
  year      = {2015},
  title     = {Understanding student motivation, behaviors and perceptions in MOOCs},
  booktitle = {Proceedings of the 18th ACM Conference on Computer Supported Cooperative Work \& Social Computing},
  pages     = {1882--1895},
  publisher = {ACM},
  doi       = {10.1145/2675133.2675217},
  url       = {https://doi.org/10.1145/2675133.2675217}
}

@misc{nasfaa2020,
  author       = {{National Association of Student Financial Aid Administrators (NASFAA)}},
  year         = {2020},
  title        = {Students face obstacles, lack of motivation in transition to remote learning amid pandemic, report finds},
  howpublished = {\url{https://www.nasfaa.org/news-item/22637/Students_Face_Obstacles_Lack_of_Motivation_in_Transition_to_Remote_Learning_Amid_Pandemic_Report_Finds}},
  note         = {Accessed: 2025-08-28},
}

@article{cannavale2025,
  author  = {Cannavale, Chiara and Claudio, Lorenza and Koroleva, Diana},
  year    = {2025},
  title   = {Digitalisation and artificial intelligence development. A cross-country analysis},
  journal = {European Journal of Innovation Management},
  volume  = {28},
  number  = {11},
  pages   = {112--130},
  date    = {2025-12-15},
  doi     = {10.1108/EJIM-07-2024-0828},
  url     = {https://doi.org/10.1108/EJIM-07-2024-0828}
}

@article{XIA2022104582,
title = {A self-determination theory (SDT) design approach for inclusive and diverse artificial intelligence (AI) education},
journal = {Computers \& Education},
volume = {189},
pages = {104582},
year = {2022},
issn = {0360-1315},
doi = {https://doi.org/10.1016/j.compedu.2022.104582},
url = {https://www.sciencedirect.com/science/article/pii/S0360131522001531},
author = {Qi Xia and Thomas K.F. Chiu and Min Lee and Ismaila Temitayo Sanusi and Yun Dai and Ching Sing Chai}
}

@inbook{inbook,
author = {Loitsch, Claudia and Striegl, Julian},
year = {2024},
month = {05},
pages = {595-612},
title = {AI for Inclusive Learning in Higher Education: Diversity, Accessibility, and Mental Health},
doi = {10.17877/DE290R-24364}
}

@misc{sun2025integratingemotionalintelligencememory,
      title={Integrating emotional intelligence, memory architecture, and gestures to achieve empathetic humanoid robot interaction in an educational setting}, 
      author={Fuze Sun and Lingyu Li and Shixiangyue Meng and Xiaoming Teng and Terry R. Payne and Paul Craig},
      year={2025},
      eprint={2505.19803},
      archivePrefix={arXiv},
      primaryClass={cs.RO},
      url={https://arxiv.org/abs/2505.19803}, 
}

@inproceedings{10.1007/978-3-031-93539-8_16,
author = {Matus, Sebastian and Cano, Sandra},
title = {Human-Robot Interaction in Higher Education: A Literature Review},
year = {2025},
isbn = {978-3-031-93538-1},
publisher = {Springer-Verlag},
address = {Berlin, Heidelberg},
url = {https://doi.org/10.1007/978-3-031-93539-8_16},
doi = {10.1007/978-3-031-93539-8_16},
booktitle = {Social Computing and Social Media: 17th International Conference, SCSM 2025, Held as Part of the 27th HCI International Conference, HCII 2025, Gothenburg, Sweden, June 22–27, 2025, Proceedings, Part I},
pages = {236–256},
numpages = {21},
location = {Gothenburg, Sweden}
}

@article{chen2023precision,
  author  = {Chen, X. and Cheng, G. and Zou, D. and Zhong, B. and Xie, H.},
  year    = {2023},
  title   = {Artificial Intelligent Robots for Precision Education: A Topic Modeling-Based Bibliometric Analysis},
  journal = {Educational Technology \& Society},
  volume  = {26},
  number  = {1},
  pages   = {171--186},
  url     = {https://www.jstor.org/stable/48707975}
}
\end{document}